\documentclass[12pt]{article}

\usepackage{arxiv}

\usepackage[utf8]{inputenc} 
\usepackage[T1]{fontenc}    
\usepackage{hyperref}       
\usepackage{url}            
\usepackage{booktabs}       
\usepackage{amsfonts}       
\usepackage{nicefrac}       
\usepackage{microtype}      
\usepackage{lipsum}
\usepackage{graphicx}
\graphicspath{ {./images/} }

\title{The Practice of Ensuring Repeatable and Reproducible Computational Models}

\author{
Herbert M Sauro \\
  Department of Bioengineering\\
  University of Washington\\
  Box 355061\\
  Seattle, WA, 98195-5061  \\
  \texttt{hsauro@uw.edu} \\
}

\begin{document}
\maketitle
\begin{abstract}
Recent studies have shown that the majority of published computational models in systems biology and physiology are not repeatable or reproducible. There are a variety of reasons for this. One of the most likely reasons is that given how busy modern researchers are and the fact that no credit is given to authors for publishing repeatable work, it is inevitable that this will be the case. The situation can only be rectified when government agencies, universities and other research institutions change policies and that journals begin to insist that published work is in fact at least repeatable if not reproducible. In this chapter guidelines are described that can be used by researchers to help make sure their work is repeatable. A scoring system is suggested that authors can use to determine how well they are doing. 
\end{abstract}


\section{Introduction}

Independent scientific confirmation has been critical to how we have built a major corpus of reliable knowledge that can be used to better the human condition whether it be for medicine, agriculture, technology or simply to better understand the world we live in. This has been understood for at least a 1000 years~\cite{Bacon:1267}. It may therefore be surprising to learn that reproducibility of scientific work has been of some recent concern~\cite{Baker2016}. Numerous articles~\cite{Fanelli2628} and reports~\cite{NAP25303} have been published that discuss the issues and possible remedies.  A recent survey by the Biomodels team~\cite{Sharif_MSB_2021} confirmed previous anecdotal evidence that a large proportion (over 50\%) of published computational models of physiological processes were essentially irreproducible. The reasons for this are still debated but one is the reward system that pervades scientific research. Rewards for promotion, hiring and tenure are often measured using metrics based on publication records and success in obtaining funding; the reproducibility of published work is not considered. With the limited bandwidth that a modern researcher has it is inevitable that aspects of their work will fall away in preference to work considered more important by  research institutions, and scientific review panels. 

In this chapter we will discuss what precautions a busy scientist can take to ensure that their work is at least repeatable. Although the focus is on computational models is systems biology, the problem of repeatable computational science is endemic~\cite{haibe2020transparency}. 

\begin{figure}
    \centering
    \includegraphics{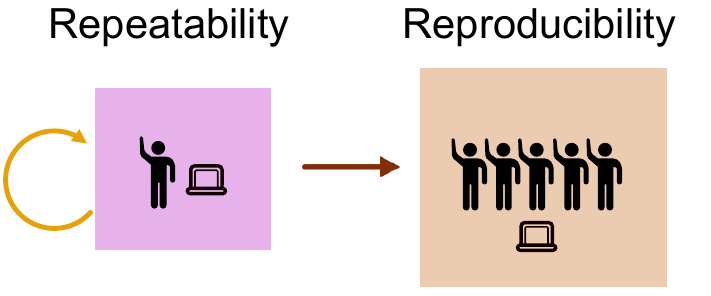}
    \caption{Repeatability and Reproducibility. In this text a computational study is repeatable when the author of the computational experiment provides all the necessary code and data to execute the computation. Reproducibility refers to the case when a third-party recreates {\em de novo} some or all of the analysis described by a researcher.}
    \label{fig:reprepro}
\end{figure}

We begin with a comment on terminology. The language used in the reproduciblity literature is quite varied and there is no single definition of what constitutes a reproducible result. There is a surprising amount of disagreement in the scientific literature on the meaning of specific terms such a reproducible, repeatable, and replicable~\cite{Barba:Terminology}. Here we will define just two terms ‘repeatability’ and ‘reproducibility’~\cite{easterbrook2014open}. These terms have been used for some time by the experimental biological communities~\cite{Blainey2014}, and their definitions closely match those used by the Association for Computing Machinery~\cite{AssociationforComputingMachinery2018}. As a personal  opinion these terms are also preferred because to {\em repeat} implies to carry out the same process exactly, while {\em reproduce} suggests creating anew (Figure~\ref{fig:reprepro}). Note that other domains have alternative definitions~\cite{Barba:Terminology,annurev-publhealth:Peng}.

Briefly, repeatability refers to a third-party researcher being able to generate the same results using the same data and computational software (preferably similar hardware and underlying operating system).  To achieve repeatability the researcher must have access to all assets that were used in the original study. In principle, the same operating system (Linux, Mac, or Windows) and computer hardware (Intel or ARM-based computers) should also be used. Often the underlying hardware and operating system are not so important but a well-publicized instance in which the underlying operating system had a marked effect on calculating NMR chemical shifts~\cite{Neupane2019} shows that even the operating system can affect the results of a computation. 

On the face of it repeating a computational experiment should be straight forward, unfortunately published studies rarely give all the necessary information in order to repeat a study. For example important documentation may be missing, required data sets absent or the code used to run the simulations is not available.

If we thought that repeating a study was difficult, reproducing a study is even more difficult. This is where one of more aspects of the publish study is recreated {\em de novo} by a third-party. This may involve writing new software, using different computational but related algorithms or even new data. It should be evident that reproducing a computational experiment and obtaining the same result is a much stronger statement that simple repeating a study. However the first stage in publishing a study is to make sure that at minimum the study is repeatable. In this article we will focus on repeatability with a discussion of reproducible modeling at the end of the article. 

In this article the main focus will be on mechanistic models published by the systems biology community. These generally include metabolic, protein signaling and gene regulatory models. These are mostly subcellular model but can include multiscale models that encompass tissues and whole organs. Model repositories such as Biomodels~\cite{LeNovere2006}, ModelDB~\cite{mcdougal2017twenty} or PMR~\cite{lloyd2008cellml} have collectively 1000s of such models for download. Previous publications have discussed specific tooling recommendations~\cite{porubsky2020best} and general surveys~\cite{porubsky2020publishing,blinov2021practical} which will not be covered here.

The most commonly used mathematical formalisms in systems biology are systems of differential equations or partial differential equations. Often the equations are be written out in full in a convenient programming language and solved using a numerical software library. One of the chief disadvantages of publishing a model in a purely mathematical form is that we lose biological context and this is one of the major limitations of publishing repeatable rather than reproducible studies. This restricts what can be done with the model post-publication, particularly reusing the model in other situations.  In that sense publishing models this way doesn't entirely follow FAIR principles~\cite{wilkinson2016fair}. 
The solution is to use representations such as SBML that are designed  to encode biological information. We will briefly discussion the use of such standards towards the end of the article.  The immediate focus of the article is what a researcher should do if they publish models in raw computational code such as Python, MATLAB etc. In general I do not recommend publishing models this way, especially large models but sometimes we must out of necessity.


\section{Repeatability: Preserving Data and Code}

Historically, a common approach to publishing a computational model in a journal was to describe the model equations in an appendix and the various parameters used in the model in one or more tables or even plots in figures. A reader was expected to transcribe the equations into a computer program, insert the values of the parameters and hope to get the same reported results. This sometimes worked, but more often than not the parameter tables were incomplete, for example missing initial conditions, missing parameter values, or the equations written out in the appendix contained one or more errors. Before the advent of the internet, this was often the only way to transmit a computational model to the community. Even today, some authors publish their models in this way. Unfortunately, unless the author takes a great deal of care, publishing a model as prose is very error prone. With the advent of code and data repositories and the move to publishing online it is now easier than ever to provide the original files that were used in the study.  

It should go without saying that repeatability and transparency in research means that publicly funded science should always release computational software as open source. An interesting example of the importance of open source software can be found in the supercooled water community~\cite{Aut2018} where the lack of software transparency resulted in over seven years of disagreement of a particular result. It  will therefore be assumed that the software is open source and freely open to inspection and reuse. 

\subsection{Scoring Repeatability}

\begin{figure}
    \centering
    \includegraphics[scale=0.4]{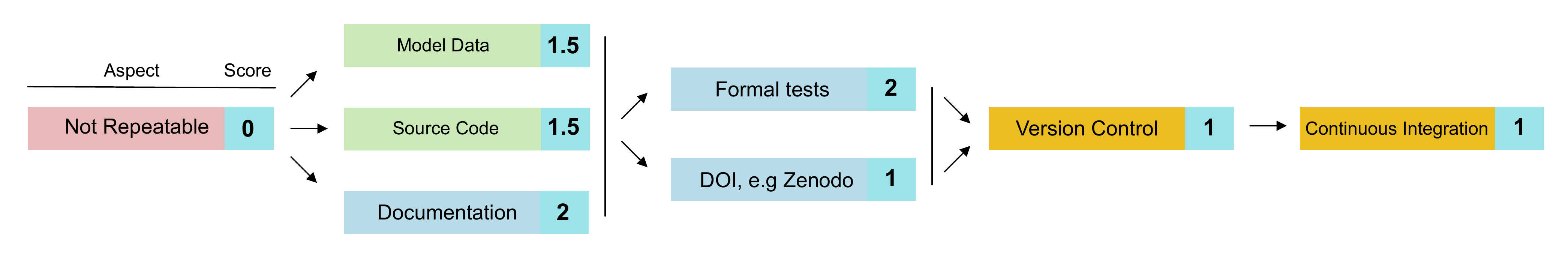}
    \caption{Suggested scoring for a repeatable study. A score of zero means the study is not-repeatable, a score of 10 is the maximum score.}
    \label{fig:levels}
\end{figure}

How can a busy researcher ensure their computational studies are repeatable? There are various degrees of quality that one can strive to ensure repeatability ranging from some basic requirements to much more sophisticated approaches involving unit testing, continuous integration and even docker images. As an example, Figure~\ref{fig:levels} suggests one possible approach to scoring quality when publishing a repeatable study, also summarised in Table~\ref{tab:compliance}. Extra weight is given to code, data and testing because these are considered the most important. Scoring is out of 10, the higher the core the better.

Other groups have developed similar criterion, most notably the Committee on Credible Practice of Modeling and Simulation in Healthcare, has developed a series of criteria for publishing computational models~\cite{mulugeta2018credibility,erdemir2020credible}. 

The lowest score of zero is reserved for published work that is irreproducible. Surprisingly even with today's emphasis on open science, examples of irreproducible results continue to be published. Three recent such examples that impacted the author's own work include a model of DNA damage~\cite{alkan2018modeling} (missing parameter sets), a model of {\em P. putida}~\cite{tokic2020large} (model not provided) and a recent model of {\em E coli}~\cite{foster2019escherichia} where a description of the model (the model itself was not available) was distributed among a large number of Excel spreadsheets making the recapitulation of the model extremely difficult. It should be emphasized that there is no malevolence or sloppiness at work here. As noted in the introduction, the pressures on a modern researcher and the current reward system will inevitably result in situations like this. However, some basic guidance on how to improve current practice will help better the situation.  

{\bf Score 0:} The published work is irreproducible. We don't ever want to be in this situation. Ideally journals should reject such submissions.
 
{\bf Code and Data (Score 3).} This is the most basic level that ensures some degree of repeatability. At this level we expect an author to supply the code that was used to run the computations and any data the software required. By code we mean that the model has been encoded into one of the common programming languages such as Python, R, MATLAB, Object Pascal, C/C++ etc. In general one should avoid proprietary languages such as MATLAB or Mathematical because not everyone has access to commercial applications plus the algorithms employed by these tools tend to be unknown. It is assumed that the code is well documented and structured, and that the data files are clearly identified with appropriate descriptive file names. Much has been written on developing good academic software~\cite{lee2018ten,romano2020ten} which readers can consult. Academia StackExchange (\url{https://academia.stackexchange.com/}) is also a good place to find discussion on the finer points of writing academic software. 

One question that arises is where to deposit the code and data? Ten years ago a common practice was to put the code on the author's personal web site and provide a link to the location in the paper. With hindsight this turned out to be a terrible idea. Such websites usually disappear along with any resources mentioned in the paper. The lesson is never put code or data on to your own personal website that is linked to a published paper. Instead there are a variety of stable open repositories that can act as hosts for code and data. The most well known of these is GitHub, which has become a popular location for academic source code. Others include Sourceforge, Bitbucket, or GitLab. These sites appear to be fairly stable. For example, sourceforge has been in existence since 1999. The only significant downside is that Git, which is used by GitHub and GitLab, has a steep learning curve. Sourceforge which uses the SVN protocol is much easier to use and is a viable option for the researcher who doesn't have the time to learn the intricacies of Git. There are also more academically orientated sites such as Zenodo~\cite{peters2017zenodo} and Dryad~\cite{white2008dryad}. Zenodo is primarily focused on the storage of software and Dryad for data. However one can store data on Zenodo as well as software. My recommendation is once a model study is ready for publication, store your modeling experiments, code and data to Zenodo. Importantly, both services offer DOIs (Digital Object Identifier, \url{https://www.doi.org/}) for deposited information. GitHub recommends that researchers archive their code to Zenodo and Zenodo offers a painless mechanism~\cite{GitHubZenodo} to automatically archive your source code from a GitHub repository. Another location where models and data can be stored is the physiome journal (\url{https://journal.physiomeproject.org/}).

What about data~\cite{hart2016ten} and what kind of data are we considering? It depends on the size of model and the analyses being applied to the model. In the simplest case it could be just tables of kinetic constants, initial conditions and possibly some validation data. For more complex situations, an author may include time-course, steady-state, or perturbation data that was used to fit the model. The current demands for data storage for kinetic models is not great compared to disciplines that use large quantities of omic data. For this reason formats for storing data can be quite simple and a variety of formats are used. Some of the most commonly used format, include basic CSV (comma separated values) files, MATLAB .mat binary files, or structured Excel based files. 

Excel is a common vehicle for storing data because of it's widespread availability, its familiarity to many researchers and the straight forward manner for doing basic data manipulation. MATLAB .mat files are a binary container format for MATLAB programs. However recent editions of MATLAB use HDF5~\cite{folk2011overview}, a community driven hierarchical data format that is supported by many software frameworks. However, given the widespread availability of Excel, the choice to use Excel is not such a bad solution. The advantage of Excel files is that the data can be more structured and Microsoft supplies a free Excel viewer for those who don't possess a copy of Excel. In addition there are other readers of Excel files such as Google Sheets. There has been some concern with using Excel for biological data due some instances of data `corruption' where for example gene name errors occurred due to Excel's confusion over date formats~\cite{zeeberg2004mistaken}. The other concern is that the Excel default file format, `.xslx' is not strictly an open standard. If in doubt store your data as CSV files or if the datasets are large and hierarchical, HDF5 is a good option. The key to data storage is to make sure that all the metadata associated with the data is clearly given. This would include items such as cell-lines, growth conditions, etc. 

If the data is much more extensive and structured, requiring considerable metadata, one might resort to using ISA-Tab~\cite{rocca2010isa}. Although originally supported via Java libraries, a recent Python interface~\cite{johnson2020isa,sansone2012toward} is likely to accelerate its uptake further. One major advantage of data stored as ISA-Tab or HDF5 is that these formats are structured in a standard way. This is unlike an Excel spreadsheet where every author will have their own way for presenting the data, making machine reading of the data difficult.  

The data purest may consider distributing data in Excel files, a grave sin, but this is far better than in some publications where data is stored in the main body of the paper and sometimes even embedded within a paragraph of text~\cite{choi2011physiological}.This is generally considered bad practice especially when models are complex.

Zenodo and Dryad have been mentioned before as places to store data, but other  notable persistent sites include figshare~\cite{Singh2011}, OpenSeek~\cite{wittig2017data,wolstencroft2015seek} and the Harvard Dataverse (\url{https://dataverse.harvard.edu/}). The work by Snoep at al~\cite{penkler2015construction}, provides a good example of the use of OpenSeek.

What's been described is the bare minimum. However, it is very easy to upload the incorrect code or to miss out an essential piece of data. As models become more complex the likelihood of this happening increases. The solution is simple however, find a friendly colleague who is willing to try to reproduce your results given the components that you provide with the manuscript. This is an easy way to identify problems, in fact the first problem they will encounter is, how do they run the code to get the results reported in the paper? This leads to a new scorable item, documentation. 

{\bf Documentation (Score 2).} The next scorable item is providing the instructions on how to run the code and generate the results. When someone has spent a year or two developing a computational model, it will seem obvious to them how someone else might run the model. To a new researcher it might not be so obvious. This is why instructional material can often be missing, or more frequently, incomplete. Documentation is therefore vital and even in the author's own lab, writing instructions for building or running software can be hard for those who are very close to the work. It is often better to have someone else write the instructions and often the best person to do this is the PI of the lab. 

Documentation is important for another reason.  Models are published in various forms for execution purposes. A common form is MATLAB or Python but authors will also use other languages such as C/C++, Java, FORTRAN, etc.  In some rare cases, a simulation may be reported using a formal workflow system such as Galaxy~\cite{Goecks2010} or Taverna~\cite{Wolstencroft2013}. An example of a Taverna workflow as applied to mechanistic modeling can be found at~\url{https://www.myexperiment.org/packs/107.html}. The publication of models is therefore fairly ad hoc. Since models are published in a wide variety of forms, repeating a study is a unique experience for almost every published model. this means good documentation is critical. 

{\bf Testing (Score 2).} Testing model code is crucial but it's not always done thoroughly or is done in a way that makes it difficult for a third-party to repeat or update.  The purpose of testing is at least two-fold. 1) To ensure that the model behaves as expected according to the original intention and 2) to make sure it much easier to refactor or add to the model, knowing that any side effects can be quickly tested and resolved. The second reason is particularly important and highlights the point that testing should be started very early in the process. In software development, most programming languages supply libraries to help with testing, for example {\tt unittest} for Python. These allow testing to be setup so that tests can be automatically run and reports delivered to the developer. 

When it comes to systems biology models there is little in the way of formal systems for testing. pySB~\cite{lopez2013programming} is however one of the few tools that has implemented testing as part of the modeling workflow but much more should be done~\cite{hellerstein2019recent}.

\begin{table}[h]
    \centering
    \begin{tabular}{llp{9cm}} \toprule
      Aspect & Score & Description \\ \midrule
      {\bf Not repeatable} & (0)  & It is not possible to repeat the results reported in the paper.  \\[5pt]
      Code and Data & (3)  &  Model code and data are provided with the publication, often as supplementary information. \\
      Documentation  & (2)  & Documentation is provided on how to use the model code and data to generate the results provided with the publication. \\ 
      Testing & (2)  & The model code is provided with a series of formal tests to verify that the code operates as designed.  \\
      DOI & (1)  & The data, model and documentation are stored on a public repository such as Zenodo, or model repositories such as Bimodels or PMR. \\ 
      Version Control & (1) & Versions of the model and data are stored using a version control system on a public repository such as GitHub, so that a history of the development of the model is kept. Particularly important for models developed by a team. \\
      Continuous Integration & (1)  & Continuous integration combined with the formal tests to ensure that any changes to the code do not result in new bugs and errors.   \\ \bottomrule \\
    \end{tabular}
    \caption{Suggested Scoring for repeatability of computational experiments.}
    \label{tab:compliance}
\end{table}

{\bf DOI (Score 1).} We've already mentioned uploading the final model and data to repositories that can provide a DOI. These include Zeondo or specialists repositories such as Biomodels and PMR.

{\bf Version Control (Score 1).}  It is inevitable that whether one is writing a research paper, textbook, a piece of software or building a computational model that in the process variants will be generated. Version control~\cite{hellerstein2019recent} is about managing those variants, more often called versions. During development, maintaining versions can be very useful since it allows one to easily go back to older copies. Of course one doesn't need specific software to do this and keeping older versions can be accomplished by the simple act of manually copying snapshots of work. In  team efforts such manual mechanisms can however be difficult to manage. Formal systems such as Git or SVN, come into play when working within a team where team members contribute to a single common code base. In software engineering this is a common practice but also works when writing multi-authored textbooks or research articles. In terms of building a model, especially when authored by a single individual, formal version control is perhaps less important and I don't believe there is a hard rule in this case. This is especially the case given that Git based version control is non-trivial to use. Use version control if its convenient. Where version control is more important in building computational models is post-publication. It is often the case that after a model is published it continues to be developed. In this situation it is essential that the model described in the original publication is preserved, and that any future changes be recorded. Using a version control system in this case is strongly encouraged in order to preserve provenance and subsequent alterations. 

{\bf Continuous Integration (Score 1).} The way professional software developers ensure repeatability is by using a technique called Continuous Integration~\cite{meyer2014continuous}. This is where software is automatically verified on a daily basis to ensure that it functions correctly. This is particularly important with a team of developers where there may be daily changes to the code-base. A series of ``regression'' tests are created which the software is expected to pass at all times. Continuous integration has not yet made any real impact in systems biology modeling but it clearly could. Some authors have commented on the use of continuous integration as a strategy to make computational work more repeatable~\cite{hellerstein2019recent,beaulieu2017reproducibility,krafczyk2019scientific}. Although perhaps more relevant during model development, continuous integration could also have significant benefits at the time of publication by providing evidence of repeatability. Continuous integration can be readily setup on GitHub with badges displayed indicating success or failure. It would be straight forward for a reviewer of a paper to inspect the badge to confirm repeatability or not.


 \subsection*{Other Distribution Options}

One can't leave the question of repeatability and distribution of modeling experiments without mentioning the use of Jupyter notebooks~\cite{perkel2018jupyter,Medley2018,badenhorst2019workflow} which have gained considerable popularity in recent years. Jupyter notebooks allow users to create live code and documentation simultaneously. This makes it very easy to create a self-contained computational experiment that can be easily distributed to others via tools such as CoLab~\cite{jeon2020setup}, Binder (\url{https://mybinder.org/}) or as a Docker image. This idea has been extended to publishing `live' papers where the published article includes embedded Jupyter like notebooks~\cite{lewis2018replication}. Creation of a live-paper is supported by the open source application Stencila (\url{https://stenci.la/}). Jupyter notebooks can also be used very effectively for teaching modeling. Exercises can be stored on Google Drive and easily used by students on CoLab. We have used this effectively to teach kinetic modeling techniques using the Python modeling package Tellurium~\cite{Choi2018,Medley2018}.

An extensive review of using technologies such as Jupyter notebook for computational research in biology was published by Hinsen~\cite{hinsen2014activepapers}. 

Another more advanced technique is the use of workflows. Workflows are often used to compose and execute a series of separate computational or data manipulation steps. Unfortunately workflows tend to be very fragile and display a trend called workflow decay~\cite{zhao2012workflows}. Over time the environment the workflow was designed to execute in has changed making it difficult or impossible to reproduce the original workflow. For example, myExperiment (\url{https://www.myexperiment.org/home}) is a hosting site for a large number of scientific workflows including a small number workflows that run biochemical simulations. However an inspection shows that the simulation workflows can no longer the executed either because the particular workflow engine is no longer supported or the resources the workflow are dependent on are no longer available. The paper by Zhao et al.~\cite{zhao2012workflows} give more detailed statistics on this issue. This makes long term use of workflows suspect. Containerizing a workflow would mitigate such issues but again it is not known how robust container technology is over time. Ideally we'd like to ensure that modeling studies, particular as they grow in size, remain viable for at least 20 to 30 years from the time the original publication was published.

In recent years~\cite{nust2020ten} a popular approach for packaging computational experiments is the Docker image and container. Docker images are very easily built so that all software libraries, applications, and other dependencies are prepared for distribution as a single entity or image. In the case of biochemical modeling workflows, relevant files would include all data used by the model, model descriptions and all simulation experiments described in the published work as we as the simulator code itself. Once an image has been created it can be distributed and run by using a Docker container. Two good examples of using this technology is the recently developed runBiosimulations resource~\cite{shaikh2021runbiosimulations} and EXSIMO. EXSIMO is an interesting case study by Matthias Konig on distributing a complete modeling experiment based, called EXSIMO~\cite{konig2020executable}. The approach packaged the data, model and code necessary to repeat the studies (\url{https://github.com/matthiaskoenig/exsimo}). The workflow used python to orchestrate all analyses using the EXSIMO python package. The model and associated data and simulation workflow were stored on a GitHub repository. This allowed model variants and simulations to be generated using GitHub's version control system. The approach at present requires some technical expertise but illustrates how a packaging systems might be implemented going forward. The example also illustrated how one might incorporate a set of tests into the workflow, akin to unit testing in software~\cite{lopez2013programming,hellerstein2019recent}. 

Docker technology can also be used to distribute applications enabling the model and its simulation studies to be manipulated and executed, through a web browser. In this case, the Docker container used to render the application, including the model, simulations and graphical output of the results, is typically hidden from the end user. While the developer will need to have expertise in generating the web application, this method ensures that end users may readily interact with a familiar GUI from the browser rather than execute commands from the command line. This approach is accessible to modelers with a range of computational backgrounds and expertise, widening the pool of potential end users who wish to repeat published simulations.

One key advantage of using Docker images is that it is guaranteed that all material to run the computational experiment is present, that is it preserves the context (or run-time environment) in which the original study was carried out.  

\section{Reproducible Computational Models}

With care, it is possible to publish repeatable computational models. Unfortunately another danger lurks in the background, something called bit-rot. This is not a new phenomenon~\cite{hayes1998bit} but is something that has plagued computing since its inception. Bit-rot refers to the situation where software that describes a model developed and published in year say 2010, no longer appears to work in 2021. The software hasn't changed but the context in which it is executed has. This is a very common, not only in modeling studies, but right across the scientific spectrum. The sad truth is that even if an author publishes a model with a score of 6, it is likely that in a few years the software may not be runnable. Container technology such as Docker can go a long way to resolve this issue but it is not known as yet how robust such approaches are over time. Anecdotal reports  (\url{https://thehftguy.com/2016/11/01/docker-in-production-an-history-of-failure/}) suggest that container technology, as a long term solution, may be limited.  


Given that the sharing of executable code is fragile and short-lived, are there examples from other domains where longer shelf-life has been achieved? Two success stories offer clues on how we might ensure repeatability and to a large extent reproduciblity in mechanistic modeling. The first example may seem frivolous but gaming consoles such as the Atari 2600~\cite{atari2600} have been available to consumers since the 1970s. Many popular games were developed for these early platforms but with the demise of the original hardware it has become increasingly difficult to reproduce the gaming experience. In other words the context in which the games were originally used has changed. The solution to this problem was the development of emulators (\url{https://en.wikipedia.org/wiki/Emulator}). These are software applications that run on modern computing hardware that mimic the original device on which the games were run. Many emulators exist for a variety of legacy platforms even for systems from the 1950s and 60s (\url{https://en.wikipedia.org/wiki/SIMH}) which are of historical importance. The key to this was the ability to recapitulate the behavior of the consoles based on the original specifications. So long as the specification is faithfully followed, it is possible to resurrect long lost gaming titles from the past. The shelf-life for these recovered games is of the order of many decades if not more. 

The second example worth mentioning is the \TeX/\LaTeX\ document preparation system developed originally by Knuth in the late 1970s and early 1980s~\cite{knuth1984texbook}. Knuth specified in detail the \TeX\ markup language as well as describing the device independent output it generated, called the DVI Driver Standard~\cite{dviStandard}. Given these specifications a number of implementations have been created that support a huge variety of output devices and formats. Today it is possible to generate output from \TeX\ files created in the 1980s even though the original software and output devices used at that time no longer exist. The shelf-life for \TeX\ documents is of the order of 40 years.

\begin{figure}
    \centering
    \includegraphics[scale=0.8]{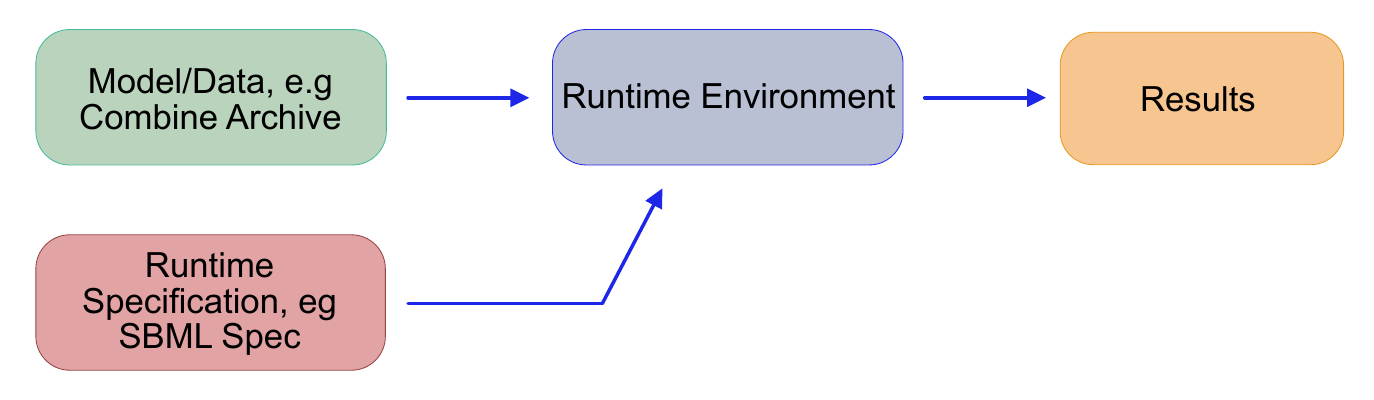}
    \caption{Reproducibility can be achieved by defining a contextual description of the problem (e.g SBML) which can be instantiated as an executable (e.g COPASI~\protect\cite{Bergmann2017}, pySB~\protect\cite{lopez2013programming}, Python, Tellurium~\protect\cite{Choi2018}, VCell~\protect\cite{Moraru}, etc) to recover the original context within which the model was be run.}
    \label{fig:runtime}
\end{figure}

These examples highlight two lessons with respect to reproducibility. The first is that there are no technical reasons to prevent successful reputability or reproduciblity~\cite{mcdougal2016reproducibility}, and second, the success of these examples was based on technical specifications that allowed uses to recreate the software environment, i.e to recover the original context (Figure~\ref{fig:runtime}). Particularly in the case of \TeX\, what is distributed is not the executable software but the \TeX\ documents that software would compile into the desired output. 

This leads us to a discussion of the Systems Biology Markup Language (SBML)~\cite{Hucka2003} and the other related standards that have emerged from the mechanistic modeling community~\cite{StandardsStatusL:2020}. SBML specifies the structure and semantics for a file format that describes a biochemical network. The SBML specification~\cite{keating2020sbml,SBMLSpec:2019} is highly detailed to ensure that any software that implements support can do so without ambiguity. Many tools have been developed to support SBML (\url{sbml.org}). Models developed in the 2000s can still be read and simulated even though the original tools that were used to create the models are no longer extant. The shelf-life for models expressed in SBML is in the order of decades. As a counter example, there are modeling papers (as described earlier) published in the just last two years whose results cannot be repeated. 

SBML spawned the development of additional standards, the most important being the Simulation Experiment Description Markup Language (SED-ML) which specifies how a simulation experiment was carried out in a published article~\cite{Waltemath2011}. SED-ML is being supported by a growing body of simulation tools~\cite{Choi2018,Medley2018}. In parallel, we also saw the development of CellML~\cite{lloyd2004cellml} which was more geared towards the exchange of physiological models. Given the variety of different standards, including modeling~\cite{Hucka2003,lloyd2004cellml}, execution~\cite{Waltemath2011}, diagramming~\cite{Novere2009,rougny2019sbgntikz} and annotation~\cite{Neal2019}, a way for packaging these separate aspects into a single file has also been devised, called the COMBINE archive~\cite{Bergmann2014}. COMBINE archives are zip files with a manifest. This file type is currently supported by a growing list of tools as well as  model repositories and is likely to be the preferred format in the future for distributing computational experiments in systems biology.  

 Once models can be formally represented it is possible to annotate models in a computer readable form. This allows additional information to be added to a model such as unambiguous naming of model participants, author and other provenance information~\cite{drager2015sbmlsqueezer,neal2019semgen}. Standards such as MIRIAM~\cite{le2005minimum} have emerged to require minimum levels of annotation in a biochemical models. Additionally the rise of exchange formats stimulated the development of new Ontologies, the most notable being, SBO (Systems Biology Ontology)~\cite{courtot2011controlled}, KiSAO (Kinetic Simulation Algorithm Ontology)~\cite{courtot2011controlled}, and OPB (Ontology of physics for biology)~\cite{cook2013ontology}. These technologies continue to be developed~\cite{Neal2019,10.1093/bioinformatics/btab445}.

The use of standards such as SBML allows some degree of reproducibility beyond repeatability. It relies on software interpreting the model correctly by compiling the specification in to a model and executing the resulting mathematical model. In addition models can be run on different simulators for verification purposes. Moreover if a given simulator is no longer available, the model can still be reproduced by using alternative simulators. This ensures that models should have a much  longer shelf-life than is currently possible when models are published are raw source code.  


\section*{Conclusion}

Publishing systems biology models that are repeatable let alone reproducible is not a glamours part of the scientific process even though it is clearly critical to the accumulation of reliable knowledge. Simple actions can be taken that can make a huge difference however the current reward  environment for researchers needs to change in order to make this more likely. One other possible avenue is for journals to demand that all published work be repeatable. Some journals are doing this. The American Journal of Political Science (AJPS) has taken the lead by using a similar approach over the last five years. They state on their website that: ``The corresponding author of a manuscript that is accepted for publication in the American Journal of Political Science must provide materials that are sufficient to enable interested researchers to verify all of the analytic results that are reported in the text and supporting materials". When AJPS receives a manuscript to review, if the manuscript is accepted for publication, it is passed to a third-party for what they term ‘verification’. The manuscript is not released to the public until verification is satisfied.  Biostatistics is another journal that has taken the lead in publishing reproducible results. This journal places kite marks (or badges) on the top-right corner of the first page of the published paper. A more detailed look at the role of badges can be found at~\cite{Nust2019}. 

PLoS Computational Biology which publishes many systems biology studies conducted a successful pilot study~\cite{papin2020improving} in 2020 in collaboration with the Center of Reproducible Biomedical Modeling (\url{https://reproduciblebiomodels.org/}) where the center acted as the body that carried out the verification of submitted models. It is likely that as more journals impose similar requirements for repeatable studies, we will see significant improvements. Another mechanism is where individual researchers take a stand by insisting that journals publish reproducible work as well as faculty candidates they review for promotion or hiring.  Such a `bottom-up' approach is advocated by Carey, et al.~{\cite{Carey2019}}. Hopefully such mechanisms can provide pressure on journals and academic institutions to begin to change their policies, and embrace the need for reproducible science.

\section*{Acknowledgements}

\section*{Funding}

This work was supported by National Institute of General Medical Sciences, National Institute for Biomedical Imaging and Bioengineering, and National Science Foundation awards R01GM123032, P41GM109824, and 1933453, respectively. The content is solely the responsibility of the authors and does not necessarily represent the official views of the National Institutes of Health, National Science Foundation, the University of Washington, or the Center for Reproducible Biomedical Modeling. 

\section*{Acknowledgements}

The author acknowledges the many fruitful discussions we have had with the COMBINE community (\url{http://co.mbine.org/}) as well as discussions with colleagues at the Center for Reproducible Biomedical Modeling~\url{https://reproduciblebiomodels.org/}.

\section*{Declaration of competing interest}

The author declares that they have no known competing financial interests or personal relationships that could have appeared to influence the work reported in this paper.

\bibliographystyle{abbrv.bst}
\bibliography{references.bib}

\begin{thebibliography}{10}

\bibitem{atari2600}
Atari 2600.
\newblock https://en.wikipedia.org/wiki/Atari\_2600, 2021.

\bibitem{GitHubZenodo}
Making your code citable.
\newblock https://guides.github.com/activities/citable-code/, 2021.

\bibitem{alkan2018modeling}
O.~Alkan, B.~Schoeberl, M.~Shah, A.~Koshkaryev, T.~Heinemann, D.~C. Drummond,
  M.~B. Yaffe, and A.~Raue.
\newblock Modeling chemotherapy-induced stress to identify rational combination
  therapies in the dna damage response pathway.
\newblock {\em Science signaling}, 11(540), 2018.

\bibitem{AssociationforComputingMachinery2018}
{Association for Computing Machinery}.
\newblock {Artifact review and badging}, 2018.

\bibitem{Aut2018}
G.~S.~A. author Aut.
\newblock {The war over supercooled water}.
\newblock {\em Physics Today}, aug 2018.

\bibitem{badenhorst2019workflow}
M.~Badenhorst, C.~J. Barry, C.~J. Swanepoel, C.~T. Van~Staden, J.~Wissing, and
  J.~M. Rohwer.
\newblock Workflow for data analysis in experimental and computational systems
  biology: Using python as ‘glue’.
\newblock {\em Processes}, 7(7):460, 2019.

\bibitem{Baker2016}
M.~Baker.
\newblock 1,500 scientists lift the lid on reproducibility.
\newblock {\em Nature}, 533(7604):452--454, may 2016.

\bibitem{Barba:Terminology}
L.~A. Barba.
\newblock Terminologies for reproducible research.
\newblock {\em CoRR}, abs/1802.03311, 2018.

\bibitem{beaulieu2017reproducibility}
B.~K. Beaulieu-Jones and C.~S. Greene.
\newblock Reproducibility of computational workflows is automated using
  continuous analysis.
\newblock {\em Nature biotechnology}, 35(4):342--346, 2017.

\bibitem{Bergmann2014}
F.~T. Bergmann, R.~Adams, S.~Moodie, J.~Cooper, M.~Glont, M.~Golebiewski,
  M.~Hucka, C.~Laibe, A.~K. Miller, D.~P. Nickerson, B.~G. Olivier,
  N.~Rodriguez, H.~M. Sauro, M.~Scharm, S.~Soiland-Reyes, D.~Waltemath,
  F.~Yvon, and N.~{Le Nov{\`{e}}re}.
\newblock {COMBINE archive and OMEX format: one file to share all information
  to reproduce a modeling project}.
\newblock {\em BMC Bioinformatics}, 15(1):369, dec 2014.

\bibitem{Bergmann2017}
F.~T. Bergmann, S.~Hoops, B.~Klahn, U.~Kummer, P.~Mendes, J.~Pahle, and
  S.~Sahle.
\newblock {COPASI and its applications in biotechnology}, nov 2017.

\bibitem{Neupane2019}
J.~Bhandari~Neupane, R.~P. Neupane, Y.~Luo, W.~Y. Yoshida, R.~Sun, and P.~G.
  Williams.
\newblock Characterization of leptazolines a–d, polar oxazolines from the
  cyanobacterium leptolyngbya sp., reveals a glitch with the
  “willoughby–hoye” scripts for calculating nmr chemical shifts.
\newblock {\em Organic Letters}, 21(20):8449--8453, 2019.
\newblock PMID: 31591889.

\bibitem{Blainey2014}
P.~Blainey, M.~Krzywinski, and N.~Altman.
\newblock Points of significance: Replication.
\newblock {\em Nature Methods}, 11(9):879--880, 9 2014.

\bibitem{blinov2021practical}
M.~L. Blinov, J.~H. Gennari, J.~R. Karr, I.~I. Moraru, D.~P. Nickerson, and
  H.~M. Sauro.
\newblock Practical resources for enhancing the reproducibility of mechanistic
  modeling in systems biology.
\newblock {\em Current Opinion in Systems Biology}, 2021.

\bibitem{Carey2019}
M.~A. Carey, A.~Dr{\"a}ger, J.~A. Papin, and J.~T. Yurkovich.
\newblock Community standards to facilitate development and address challenges
  in metabolic modeling.
\newblock {\em bioRxiv}, page 700112, 2019.

\bibitem{choi2011physiological}
J.-S. Choi and S.~G. Waxman.
\newblock Physiological interactions between nav1. 7 and nav1. 8 sodium
  channels: a computer simulation study.
\newblock {\em Journal of neurophysiology}, 106(6):3173--3184, 2011.

\bibitem{Choi2018}
K.~Choi, J.~K. Medley, M.~K{\"{o}}nig, K.~Stocking, L.~Smith, S.~Gu, and H.~M.
  Sauro.
\newblock {Tellurium: An extensible python-based modeling environment for
  systems and synthetic biology}.
\newblock {\em Biosystems}, 171:74--79, sep 2018.

\bibitem{cook2013ontology}
D.~L. Cook, M.~L. Neal, F.~L. Bookstein, and J.~H. Gennari.
\newblock Ontology of physics for biology: representing physical dependencies
  as a basis for biological processes.
\newblock {\em Journal of biomedical semantics}, 4(1):1--8, 2013.

\bibitem{courtot2011controlled}
M.~Courtot, N.~Juty, C.~Kn{\"u}pfer, D.~Waltemath, A.~Zhukova, A.~Dr{\"a}ger,
  M.~Dumontier, A.~Finney, M.~Golebiewski, J.~Hastings, et~al.
\newblock Controlled vocabularies and semantics in systems biology.
\newblock {\em Molecular systems biology}, 7(1):543, 2011.

\bibitem{drager2015sbmlsqueezer}
A.~Dr{\"a}ger, D.~C. Zielinski, R.~Keller, M.~Rall, J.~Eichner, B.~O. Palsson,
  and A.~Zell.
\newblock Sbmlsqueezer 2: context-sensitive creation of kinetic equations in
  biochemical networks.
\newblock {\em BMC systems biology}, 9(1):1--17, 2015.

\bibitem{easterbrook2014open}
S.~M. Easterbrook.
\newblock Open code for open science?
\newblock {\em Nature Geoscience}, 7(11):779, 2014.

\bibitem{erdemir2020credible}
A.~Erdemir, L.~Mulugeta, J.~P. Ku, A.~Drach, M.~Horner, T.~M. Morrison, G.~C.
  Peng, R.~Vadigepalli, W.~W. Lytton, and J.~G. Myers.
\newblock Credible practice of modeling and simulation in healthcare: ten rules
  from a multidisciplinary perspective.
\newblock {\em Journal of translational medicine}, 18(1):1--18, 2020.

\bibitem{Fanelli2628}
D.~Fanelli.
\newblock Opinion: Is science really facing a reproducibility crisis, and do we
  need it to?
\newblock {\em Proceedings of the National Academy of Sciences},
  115(11):2628--2631, 2018.

\bibitem{folk2011overview}
M.~Folk, G.~Heber, Q.~Koziol, E.~Pourmal, and D.~Robinson.
\newblock An overview of the hdf5 technology suite and its applications.
\newblock In {\em Proceedings of the EDBT/ICDT 2011 Workshop on Array
  Databases}, pages 36--47, 2011.

\bibitem{foster2019escherichia}
C.~J. Foster, S.~Gopalakrishnan, M.~R. Antoniewicz, and C.~D. Maranas.
\newblock From escherichia coli mutant 13c labeling data to a core kinetic
  model: A kinetic model parameterization pipeline.
\newblock {\em PLoS computational biology}, 15(9):e1007319, 2019.

\bibitem{Goecks2010}
J.~Goecks, A.~Nekrutenko, J.~Taylor, and T.~{Galaxy Team}.
\newblock {Galaxy: a comprehensive approach for supporting accessible,
  reproducible, and transparent computational research in the life sciences}.
\newblock {\em Genome Biology}, 11(8):R86, 2010.

\bibitem{haibe2020transparency}
B.~Haibe-Kains, G.~A. Adam, A.~Hosny, F.~Khodakarami, L.~Waldron, B.~Wang,
  C.~McIntosh, A.~Goldenberg, A.~Kundaje, C.~S. Greene, et~al.
\newblock Transparency and reproducibility in artificial intelligence.
\newblock {\em Nature}, 586(7829):E14--E16, 2020.

\bibitem{hart2016ten}
E.~M. Hart, P.~Barmby, D.~LeBauer, F.~Michonneau, S.~Mount, P.~Mulrooney,
  T.~Poisot, K.~H. Woo, N.~B. Zimmerman, and J.~W. Hollister.
\newblock Ten simple rules for digital data storage, 2016.

\bibitem{hayes1998bit}
B.~Hayes.
\newblock Bit rot.
\newblock {\em American Scientist}, 86(5):410--415, 1998.

\bibitem{hellerstein2019recent}
J.~L. Hellerstein, S.~Gu, K.~Choi, and H.~M. Sauro.
\newblock Recent advances in biomedical simulations: a manifesto for model
  engineering.
\newblock {\em F1000Research}, 8, 2019.

\bibitem{hinsen2014activepapers}
K.~Hinsen.
\newblock Activepapers: a platform for publishing and archiving computer-aided
  research.
\newblock {\em F1000Research}, 3, 2014.

\bibitem{SBMLSpec:2019}
M.~Hucka, F.~T. Bergmann, C.~Chaouiya, A.~Dräger, S.~Hoops, S.~M. Keating,
  M.~König, N.~L. Novère, C.~J. Myers, B.~G. Olivier, S.~Sahle, J.~C. Schaff,
  R.~Sheriff, L.~P. Smith, D.~Waltemath, D.~J. Wilkinson, and F.~Zhang.
\newblock The systems biology markup language (sbml): Language specification
  for level 3 version 2 core release 2.
\newblock {\em Journal of Integrative Bioinformatics}, 16(2), 2019.

\bibitem{Hucka2003}
M.~Hucka, A.~Finney, H.~M. Sauro, H.~Bolouri, J.~C. Doyle, H.~Kitano, A.~P.
  {and the rest of the SBML Forum:}, A.~P. Arkin, B.~J. Bornstein, D.~Bray,
  A.~Cornish-Bowden, A.~A. Cuellar, S.~Dronov, E.~D. Gilles, M.~Ginkel, V.~Gor,
  I.~I. Goryanin, W.~J. Hedley, T.~C. Hodgman, J.-H. Hofmeyr, P.~J. Hunter,
  N.~S. Juty, J.~L. Kasberger, A.~Kremling, U.~Kummer, N.~{Le Novere}, L.~M.
  Loew, D.~Lucio, P.~Mendes, E.~Minch, E.~D. Mjolsness, Y.~Nakayama, M.~R.
  Nelson, P.~F. Nielsen, T.~Sakurada, J.~C. Schaff, B.~E. Shapiro, T.~S.
  Shimizu, H.~D. Spence, J.~Stelling, K.~Takahashi, M.~Tomita, J.~Wagner, and
  J.~Wang.
\newblock {The systems biology markup language (SBML): a medium for
  representation and exchange of biochemical network models}.
\newblock {\em Bioinformatics}, 19(4):524--531, mar 2003.

\bibitem{jeon2020setup}
J.~Jeon and H.~U. Kim.
\newblock Setup of a scientific computing environment for computational
  biology: Simulation of a genome-scale metabolic model of escherichia coli as
  an example.
\newblock {\em Journal of Microbiology}, 58(3):227--234, 2020.

\bibitem{johnson2020isa}
D.~Johnson, K.~Cochrane, R.~P. Davey, A.~Etuk, A.~Gonzalez-Beltran, K.~Haug,
  M.~Izzo, M.~Larralde, T.~N. Lawson, A.~Minotto, et~al.
\newblock Isa api: An open platform for interoperable life science experimental
  metadata.
\newblock {\em bioRxiv}, 2020.

\bibitem{keating2020sbml}
S.~M. Keating, D.~Waltemath, M.~K{\"o}nig, F.~Zhang, A.~Dr{\"a}ger,
  C.~Chaouiya, F.~T. Bergmann, A.~Finney, C.~S. Gillespie, T.~Helikar, et~al.
\newblock Sbml level 3: an extensible format for the exchange and reuse of
  biological models.
\newblock {\em Molecular systems biology}, 16(8):e9110, 2020.

\bibitem{knuth1984texbook}
D.~Knuth.
\newblock {\em TeXbook}.
\newblock Addison-Wesley Reading, 1984.

\bibitem{konig2020executable}
M.~K{\"o}nig.
\newblock Executable simulation model of the liver.
\newblock {\em bioRxiv}, 2020.

\bibitem{krafczyk2019scientific}
M.~Krafczyk, A.~Shi, A.~Bhaskar, D.~Marinov, and V.~Stodden.
\newblock Scientific tests and continuous integration strategies to enhance
  reproducibility in the scientific software context.
\newblock In {\em Proceedings of the 2nd International Workshop on Practical
  Reproducible Evaluation of Computer Systems}, pages 23--28, 2019.

\bibitem{LeNovere2006}
N.~{Le Novere}.
\newblock {BioModels Database: a free, centralized database of curated,
  published, quantitative kinetic models of biochemical and cellular systems}.
\newblock {\em Nucleic Acids Research}, 34(90001):D689--D691, jan 2006.

\bibitem{le2005minimum}
N.~Le~Nov{\`e}re, A.~Finney, M.~Hucka, U.~S. Bhalla, F.~Campagne,
  J.~Collado-Vides, E.~J. Crampin, M.~Halstead, E.~Klipp, P.~Mendes, et~al.
\newblock Minimum information requested in the annotation of biochemical models
  (miriam).
\newblock {\em Nature biotechnology}, 23(12):1509--1515, 2005.

\bibitem{lee2018ten}
B.~D. Lee.
\newblock Ten simple rules for documenting scientific software, 2018.

\bibitem{lewis2018replication}
L.~M. Lewis, M.~C. Edwards, Z.~R. Meyers, C.~C. Talbot~Jr, H.~Hao, D.~Blum,
  et~al.
\newblock Replication study: Transcriptional amplification in tumor cells with
  elevated c-myc.
\newblock {\em Elife}, 7:e30274, 2018.

\bibitem{lloyd2004cellml}
C.~M. Lloyd, M.~D. Halstead, and P.~F. Nielsen.
\newblock Cellml: its future, present and past.
\newblock {\em Progress in biophysics and molecular biology}, 85(2-3):433--450,
  2004.

\bibitem{lloyd2008cellml}
C.~M. Lloyd, J.~R. Lawson, P.~J. Hunter, and P.~F. Nielsen.
\newblock The cellml model repository.
\newblock {\em Bioinformatics}, 24(18):2122--2123, 2008.

\bibitem{lopez2013programming}
C.~F. Lopez, J.~L. Muhlich, J.~A. Bachman, and P.~K. Sorger.
\newblock Programming biological models in python using pysb.
\newblock {\em Molecular systems biology}, 9(1):646, 2013.

\bibitem{mcdougal2016reproducibility}
R.~A. McDougal, A.~S. Bulanova, and W.~W. Lytton.
\newblock Reproducibility in computational neuroscience models and simulations.
\newblock {\em IEEE Transactions on Biomedical Engineering}, 63(10):2021--2035,
  2016.

\bibitem{mcdougal2017twenty}
R.~A. McDougal, T.~M. Morse, T.~Carnevale, L.~Marenco, R.~Wang, M.~Migliore,
  P.~L. Miller, G.~M. Shepherd, and M.~L. Hines.
\newblock Twenty years of modeldb and beyond: building essential modeling tools
  for the future of neuroscience.
\newblock {\em Journal of computational neuroscience}, 42(1):1--10, 2017.

\bibitem{Medley2018}
J.~K. Medley, K.~Choi, M.~K{\"{o}}nig, L.~Smith, S.~Gu, J.~Hellerstein, S.~C.
  Sealfon, and H.~M. Sauro.
\newblock {Tellurium notebooks—An environment for reproducible dynamical
  modeling in systems biology}.
\newblock {\em PLOS Computational Biology}, 14(6):e1006220, jun 2018.

\bibitem{meyer2014continuous}
M.~Meyer.
\newblock Continuous integration and its tools.
\newblock {\em IEEE software}, 31(3):14--16, 2014.

\bibitem{Moraru}
I.~I. Moraru, J.~C. Schaff, B.~M. Slepchenko, M.~Blinov, F.~Morgan,
  A.~Lakshminarayana, F.~Gao, Y.~Li, and L.~M. Loew.
\newblock {The Virtual Cell Modeling and Simulation Software Environment}.

\bibitem{mulugeta2018credibility}
L.~Mulugeta, A.~Drach, A.~Erdemir, C.~A. Hunt, M.~Horner, J.~P. Ku, J.~G.
  Myers~Jr, R.~Vadigepalli, and W.~W. Lytton.
\newblock Credibility, replicability, and reproducibility in simulation for
  biomedicine and clinical applications in neuroscience.
\newblock {\em Frontiers in neuroinformatics}, 12:18, 2018.

\bibitem{NAP25303}
{National Academies of Sciences, Engineering, and Medicine}.
\newblock {\em Reproducibility and Replicability in Science}.
\newblock The National Academies Press, Washington, DC, 2019.

\bibitem{Neal2019}
M.~L. Neal, M.~K{\"{o}}nig, D.~Nickerson, G.~Misirli, R.~Kalbasi,
  A.~Dr{\"{a}}ger, K.~Atalag, V.~Chelliah, M.~T. Cooling, D.~L. Cook, S.~Crook,
  M.~{De Alba}, S.~H. Friedman, A.~Garny, J.~H. Gennari, P.~Gleeson,
  M.~Golebiewski, M.~Hucka, N.~Juty, C.~Myers, B.~G. Olivier, H.~M. Sauro,
  M.~Scharm, J.~L. Snoep, V.~Tour{\'{e}}, A.~Wipat, O.~Wolkenhauer, and
  D.~Waltemath.
\newblock {Harmonizing semantic annotations for computational models in
  biology}.
\newblock {\em Briefings in Bioinformatics}, 20(2):540--550, mar 2019.

\bibitem{neal2019semgen}
M.~L. Neal, C.~T. Thompson, K.~G. Kim, R.~C. James, D.~L. Cook, B.~E. Carlson,
  and J.~H. Gennari.
\newblock Semgen: a tool for semantics-based annotation and composition of
  biosimulation models.
\newblock {\em Bioinformatics}, 35(9):1600--1602, 2019.

\bibitem{Novere2009}
N.~L. Nov{\`{e}}re, M.~Hucka, H.~Mi, S.~Moodie, F.~Schreiber, A.~Sorokin,
  E.~Demir, K.~Wegner, M.~I. Aladjem, S.~M. Wimalaratne, F.~T. Bergman,
  R.~Gauges, P.~Ghazal, H.~Kawaji, L.~Li, Y.~Matsuoka, A.~Vill{\'{e}}ger, S.~E.
  Boyd, L.~Calzone, M.~Courtot, U.~Dogrusoz, T.~C. Freeman, A.~Funahashi,
  S.~Ghosh, A.~Jouraku, S.~Kim, F.~Kolpakov, A.~Luna, S.~Sahle, E.~Schmidt,
  S.~Watterson, G.~Wu, I.~Goryanin, D.~B. Kell, C.~Sander, H.~Sauro, J.~L.
  Snoep, K.~Kohn, and H.~Kitano.
\newblock {The systems biology graphical notation}.
\newblock {\em Nature Biotechnology}, 27(8):735--741, aug 2009.

\bibitem{Nust2019}
D.~N{\"{u}}st, L.~Lohoff, L.~Einfeldt, N.~Gavish, M.~G{\"{o}}tza, S.~T. Jaswal,
  S.~Khalid, L.~Meierkort, M.~Mohr, C.~Rendel, and A.~van Eek.
\newblock {Guerrilla Badges for Reproducible Geospatial Data Science (AGILE
  2019 Short Paper)}.
\newblock {\em AGILE 2019}, 2019.

\bibitem{nust2020ten}
D.~N{\"u}st, V.~Sochat, B.~Marwick, S.~J. Eglen, T.~Head, T.~Hirst, and B.~D.
  Evans.
\newblock Ten simple rules for writing dockerfiles for reproducible data
  science, 2020.

\bibitem{papin2020improving}
J.~A. Papin, F.~Mac~Gabhann, H.~M. Sauro, D.~Nickerson, and A.~Rampadarath.
\newblock Improving reproducibility in computational biology research, 2020.

\bibitem{annurev-publhealth:Peng}
R.~D. Peng and S.~C. Hicks.
\newblock Reproducible research: A retrospective.
\newblock {\em Annual Review of Public Health}, 42(1):null, 2021.
\newblock PMID: 33467923.

\bibitem{penkler2015construction}
G.~Penkler, F.~Du~Toit, W.~Adams, M.~Rautenbach, D.~C. Palm, D.~D. Van~Niekerk,
  and J.~L. Snoep.
\newblock Construction and validation of a detailed kinetic model of glycolysis
  in plasmodium falciparum.
\newblock {\em The FEBS journal}, 282(8):1481--1511, 2015.

\bibitem{perkel2018jupyter}
J.~M. Perkel.
\newblock Why jupyter is data scientists' computational notebook of choice.
\newblock {\em Nature}, 563(7732):145--147, 2018.

\bibitem{peters2017zenodo}
I.~Peters, P.~Kraker, E.~Lex, C.~Gumpenberger, and J.~I. Gorraiz.
\newblock Zenodo in the spotlight of traditional and new metrics.
\newblock {\em Frontiers in Research Metrics and Analytics}, 2:13, 2017.

\bibitem{porubsky2020publishing}
V.~Porubsky, L.~Smith, and H.~M. Sauro.
\newblock Publishing reproducible dynamic kinetic models.
\newblock {\em Briefings in Bioinformatics}, 2020.

\bibitem{porubsky2020best}
V.~L. Porubsky, A.~P. Goldberg, A.~K. Rampadarath, D.~P. Nickerson, J.~R. Karr,
  and H.~M. Sauro.
\newblock Best practices for making reproducible biochemical models.
\newblock {\em Cell Systems}, 11(2):109--120, 2020.

\bibitem{rocca2010isa}
P.~Rocca-Serra, M.~Brandizi, E.~Maguire, N.~Sklyar, C.~Taylor, K.~Begley,
  D.~Field, S.~Harris, W.~Hide, O.~Hofmann, et~al.
\newblock Isa software suite: supporting standards-compliant experimental
  annotation and enabling curation at the community level.
\newblock {\em Bioinformatics}, 26(18):2354--2356, 2010.

\bibitem{Bacon:1267}
B.~Roger.
\newblock {\em Opus Majus}.
\newblock 1267.
\newblock Translation: Bacon, R., 2016, Part 1. Opus Majus, Volumes 1 and 2.
  University of Pennsylvania Press, SN - 9781512814064.

\bibitem{romano2020ten}
J.~D. Romano and J.~H. Moore.
\newblock Ten simple rules for writing a paper about scientific software, 2020.

\bibitem{rougny2019sbgntikz}
A.~Rougny.
\newblock sbgntikz—a ti k z library to draw sbgn maps.
\newblock {\em Bioinformatics}, 35(21):4499--4500, 2019.

\bibitem{sansone2012toward}
S.-A. Sansone, P.~Rocca-Serra, D.~Field, E.~Maguire, C.~Taylor, O.~Hofmann,
  H.~Fang, S.~Neumann, W.~Tong, L.~Amaral-Zettler, et~al.
\newblock Toward interoperable bioscience data.
\newblock {\em Nature genetics}, 44(2):121--126, 2012.

\bibitem{StandardsStatusL:2020}
F.~Schreiber, B.~Sommer, T.~Czauderna, M.~Golebiewski, T.~E. Gorochowski,
  M.~Hucka, S.~M. Keating, M.~König, C.~Myers, D.~Nickerson, and D.~Waltemath.
\newblock Specifications of standards in systems and synthetic biology: status
  and developments in 2020.
\newblock {\em Journal of Integrative Bioinformatics}, 17(2-3), 2020.

\bibitem{shaikh2021runbiosimulations}
B.~Shaikh, G.~Marupilla, M.~Wilson, M.~L. Blinov, I.~I. Moraru, and J.~R. Karr.
\newblock runbiosimulations: an extensible web application that simulates a
  wide range of computational modeling frameworks, algorithms, and formats.
\newblock {\em bioRxiv}, 2021.

\bibitem{Singh2011}
J.~Singh.
\newblock {FigShare}, apr 2011.

\bibitem{dviStandard}
{The TUG DVI Driver Standards Committee}.
\newblock The dvi driver standard, level 0.

\bibitem{Sharif_MSB_2021}
K.~Tiwari, S.~Kananathan, M.~G. Roberts, J.~P. Meyer, M.~U. Sharif~Shohan,
  A.~Xavier, M.~Maire, A.~Zyoud, J.~Men, S.~Ng, T.~V.~N. Nguyen, M.~Glont,
  H.~Hermjakob, and R.~S. Malik-Sheriff.
\newblock Reproducibility in systems biology modelling.
\newblock {\em Molecular Systems Biology}, 17(2):e9982, 2021.

\bibitem{tokic2020large}
M.~Tokic, V.~Hatzimanikatis, and L.~Miskovic.
\newblock Large-scale kinetic metabolic models of pseudomonas putida kt2440 for
  consistent design of metabolic engineering strategies.
\newblock {\em Biotechnology for biofuels}, 13(1):1--19, 2020.

\bibitem{Waltemath2011}
D.~Waltemath, R.~Adams, F.~T. Bergmann, M.~Hucka, F.~Kolpakov, A.~K. Miller,
  I.~I. Moraru, D.~Nickerson, S.~Sahle, J.~L. Snoep, and N.~{Le Nov{\`{e}}re}.
\newblock {Reproducible computational biology experiments with SED-ML - The
  Simulation Experiment Description Markup Language}.
\newblock {\em BMC Systems Biology}, 5(1):198, 2011.

\bibitem{10.1093/bioinformatics/btab445}
C.~Welsh, D.~P. Nickerson, A.~Rampadarath, M.~L. Neal, H.~M. Sauro, and J.~H.
  Gennari.
\newblock {libOmexMeta: Enabling semantic annotation of models to support FAIR
  principles}.
\newblock {\em Bioinformatics}, 06 2021.
\newblock btab445.

\bibitem{white2008dryad}
H.~White, S.~Carrier, A.~Thompson, J.~Greenberg, and R.~Scherle.
\newblock The dryad data repository: A singapore framework metadata
  architecture in a dspace environment.
\newblock In {\em Dublin core conference}, pages 157--162, 2008.

\bibitem{wilkinson2016fair}
M.~D. Wilkinson, M.~Dumontier, I.~J. Aalbersberg, G.~Appleton, M.~Axton,
  A.~Baak, N.~Blomberg, J.-W. Boiten, L.~B. da~Silva~Santos, P.~E. Bourne,
  et~al.
\newblock The fair guiding principles for scientific data management and
  stewardship.
\newblock {\em Scientific data}, 3, 2016.

\bibitem{wittig2017data}
U.~Wittig, M.~Rey, A.~Weidemann, and W.~M{\"u}ller.
\newblock Data management and data enrichment for systems biology projects.
\newblock {\em Journal of biotechnology}, 261:229--237, 2017.

\bibitem{Wolstencroft2013}
K.~Wolstencroft, R.~Haines, D.~Fellows, A.~Williams, D.~Withers, S.~Owen,
  S.~Soiland-Reyes, I.~Dunlop, A.~Nenadic, P.~Fisher, J.~Bhagat, K.~Belhajjame,
  F.~Bacall, A.~Hardisty, A.~{Nieva de la Hidalga}, M.~P. {Balcazar Vargas},
  S.~Sufi, and C.~Goble.
\newblock {The Taverna workflow suite: designing and executing workflows of Web
  Services on the desktop, web or in the cloud.}
\newblock {\em Nucleic acids research}, 41(Web Server issue), 2013.

\bibitem{wolstencroft2015seek}
K.~Wolstencroft, S.~Owen, O.~Krebs, Q.~Nguyen, N.~J. Stanford, M.~Golebiewski,
  A.~Weidemann, M.~Bittkowski, L.~An, D.~Shockley, et~al.
\newblock Seek: a systems biology data and model management platform.
\newblock {\em BMC systems biology}, 9(1):1--12, 2015.

\bibitem{zeeberg2004mistaken}
B.~R. Zeeberg, J.~Riss, D.~W. Kane, K.~J. Bussey, E.~Uchio, W.~M. Linehan,
  J.~C. Barrett, and J.~N. Weinstein.
\newblock Mistaken identifiers: gene name errors can be introduced
  inadvertently when using excel in bioinformatics.
\newblock {\em BMC bioinformatics}, 5(1):1--6, 2004.

\bibitem{zhao2012workflows}
J.~Zhao, J.~M. Gomez-Perez, K.~Belhajjame, G.~Klyne, E.~Garcia-Cuesta,
  A.~Garrido, K.~Hettne, M.~Roos, D.~De~Roure, and C.~Goble.
\newblock Why workflows break—understanding and combating decay in taverna
  workflows.
\newblock In {\em 2012 ieee 8th international conference on e-science}, pages
  1--9. IEEE, 2012.

\end{thebibliography}

\end{document}